\RequirePackage{lineno}
\documentclass[aps,prl,twocolumn,showpacs,groupedaddress]{revtex4-1}  % for review and submission
\usepackage{graphicx}  % needed for figures
\usepackage{dcolumn}   % needed for some tables
\usepackage{bm}        % for math
\usepackage{amssymb}   % for math
\usepackage{epstopdf} %to enable use of .eps files with pdfLatex
\usepackage{amsmath}
\usepackage{url}
\newcommand{\bra}{\langle}
\newcommand{\ket}{\rangle}
\newcommand{\rkq}{\rho_q^{(\kappa)}} %apm is atomic polarization moment

\begin{document}
%% line numbers
%\setpagewiselinenumbers
%\modulolinenumbers[5]

% The following information is for internal review, please remove them for submission
%\leftline{Version xx as of \today}
%\leftline{Primary authors: Joe E. Physics}
%\leftline{To be submitted to (PRL, PRD-RC, PRD, PLB; choose one.)}
%\rightline{\em D\O\ INTERNAL DOCUMENT -- NOT FOR PUBLIC DISTRIBUTION}
%\rightline{Comment to {\tt d0-run2eb-nnn@fnal.gov}}
%\rightline{by xxx, yyy}

% the following line is for submission, including submission to the arXiv!!
%\hspace{5.2in} \mbox{Fermilab-Pub-04/xxx-E}

\title{Magic frequencies in atom-light interaction for precision probing of the density matrix}
\author{Menachem Givon,$^{1}$ Yair Margalit,$^{1}$ Amir Waxman,$^{1}$ Tal David$^{2}$,  David Groswasser,$^{1}$ Yonathan Japha$ ^{1}$ and Ron Folman$^{1}$}
\affiliation{$^{1}$Ben-Gurion University of the Negev, P.O.B. 653, Beer-Sheva 84105, Israel\\
$^{2}$Israel Aerospace Industries, Ramta Division, 1 Nafha St., Beer Sheva 84102, Israel}
\date{\today}

\begin{abstract}
We analyze theoretically and experimentally the existence of a {\it magic frequency} for which the
absorption of a linearly polarized light beam by vapor alkali atoms is independent of the population distribution
among the Zeeman sub-levels and the angle between the beam and a magnetic field. The phenomenon
originates from a peculiar cancelation of the contributions of higher moments of the atomic density matrix,
and is described using the Wigner-Eckart theorem and inherent properties of Clebsch-Gordan coefficients.
One important application is the robust measurement of the hyperfine population.
\end{abstract}

\maketitle

Interaction of light with alkali metal vapor has an important role both in the study of fundamental physics and in many technological applications. Macroscopic entanglement was demonstrated using cesium vapor cells \cite{julsgaard2001experimental}; Pulses of light, as well as images, were stored in rubidium vapor \cite{phillips2001storage,shuker2008storing}; Rubidium vapor cells serve as a basis for chip-scale atomic clocks \cite{knappe2004microfabricated}; and alkali vapor cells are used for high sensitivity optical magnetometry \cite{kominis2003subfemtotesla,budker2007optical}. More applications can be found in \cite{wieman1976doppler, demtroder2003laser, kitching2000microwave, auzinsh2002optically}.

The density matrix representing the state of an alkali vapor in a specific hyperfine state $|F\rangle$ can be
expanded in terms of polarization moments (PM). A PM $\rho^{(\kappa)}$ is an irreducible spherical
tensor of rank $\kappa$ ($0\leq \kappa\leq 2F+1$) whose components are given by
\cite{auzinsh2002optically}:
\begin{equation}
\rkq=\sum_{m_1,m_2}(-1)^{F-m_1}\bra F,m_2,F,-m_1|\kappa, q\ket\rho_{m_1,m_2}
\label{Eq:PM}
\end{equation}
where $q=-\kappa...\kappa$, $m_1$ and $m_2$ are the magnetic quantum numbers, $\rho_{m_1,m_2}$
are the density matrix elements and $\bra F,m_2,F,-m_1|\kappa, q\ket$ are the Clebsch-Gordan
coefficients (CGC). The PMs of ranks $\kappa=0,1$ and $2$  are proportional to the population, the dipole
moment (or orientation), and the quadrupole moment (or alignment), of the relevant $|F\rangle$ state,
respectively \cite{auzinsh2002optically, budker2002resonant, yashchuk2003selective, graf2005relaxation}.
For a full mapping of the density matrix it is essential to measure the hyperfine population $\rho^{(0)}$. While unpolarized light from lamps \cite{Bellbloom1958,Arditi1964,BOU1966} or polarized light from lasers \cite{Rahman1987} were previously used to optically pump atomic vapor and monitor its relaxation process, these measurements had limited accuracy in estimating the hyperfine population.

Here we demonstrate for the first time the existence of a {\it magic frequency} for which the
light absorption of linearly polarized laser radiation is independent of the population distribution among the Zeeman sub-levels and of the angles between the light and the magnetic field. At the {\it magic frequency} the absorption is proportional only to $\rho^{(0)}$ and can therefore serve as an accurate and robust measure of the hyperfine population.

\begin{figure}[t!]\centering \includegraphics[width=0.49\textwidth]{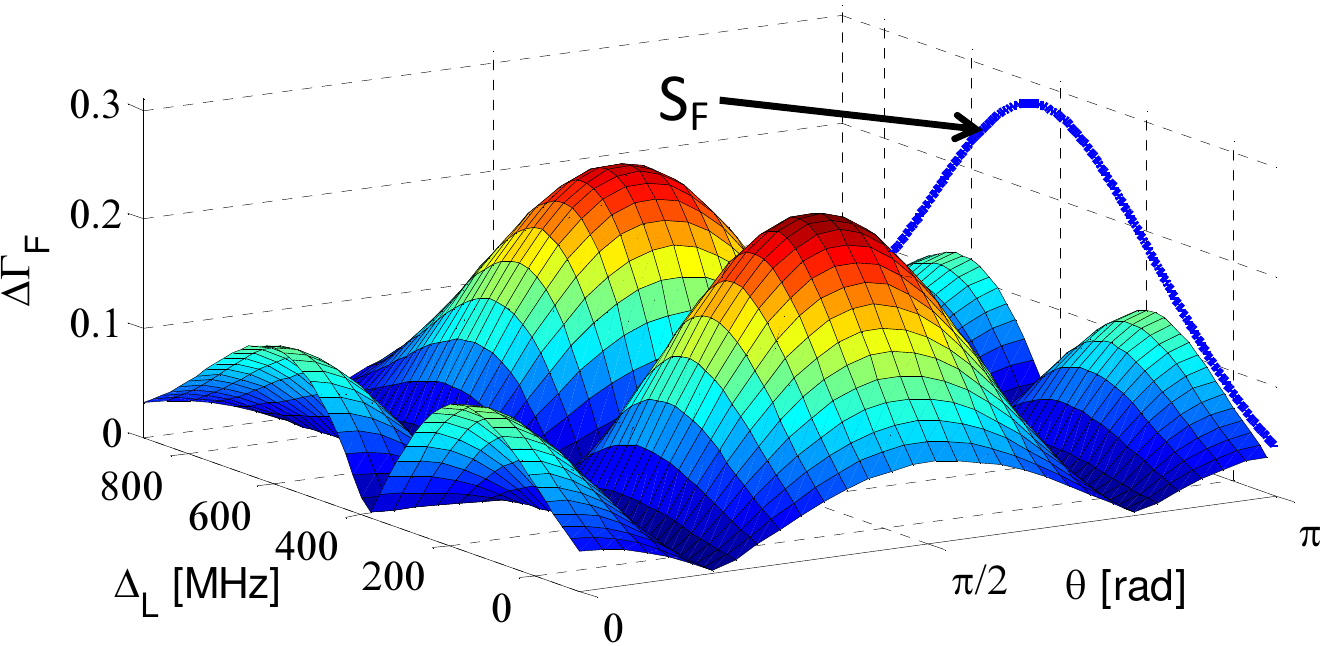}
\caption{(Color online)
Outline of the {\it magic frequency} effect. At these frequencies, the absorption from all Zeeman
sub-levels is equal, namely, the overall absorption depends only on the total population in the hyperfine state, and is independent of the internal population distribution.
Plotted is a $\Delta\Gamma_F$ surface,
proportional to the differences of the absorption rates of light ($D_2$ line) by $^{87}$Rb atoms ($T=295 K$) in different Zeeman sub-levels of the $|F=2\rangle$ hyperfine state [Eq. (\ref {Eq:DeltaGamma})], for a small external DC magnetic field ${|\bf B|}<1$ G.
The {\it magic frequency} where $\Delta\Gamma_F=0$, is found at $\Delta_L^{Magic}=385$ MHz (and at $-318$, not shown), where $\Delta_L$ is the light frequency detuning from the $\vert F=2\rangle$ to $\vert F'=0\rangle$ energy difference.
The light, propagating in direction ${\bf k}$, is linearly polarized with an angle $\phi=0$ (i.e. polarized in the ${\bf B}$-${\bf k}$ plane).
The plot also shows that the {\it magic frequency} is independent of $\theta$, the angle between $\mathbf{B}$ and $\mathbf{k}$.
The overall transition strength $S_F$ [Eq. (\ref{Eq:SF})] is shown in the background. For clarity, $\Delta\Gamma_F$ is multiplied by 3.} \label{fig:3d} \end{figure}

In general, different Zeeman sub-levels $|F,m_F\rangle$ have different contributions to the absorption rate, due to different optical transition matrix elements. To find conditions that nullify these differences, we define the value $\Delta\Gamma_F$, which measures the difference in the absorption rates from different Zeeman sub-levels in a specific hyperfine level $|F\rangle$, and which depends on the probe frequency, its direction, polarization and the vapor temperature. The model produces surfaces of $\Delta\Gamma_F$  (Fig. \ref{fig:3d}), showing if and where  $\Delta\Gamma_F=0$, thereby identifying the {\it magic frequency}.

Our model is based on the Wigner-Eckart theorem \cite{ auzinsh2002optically}. Working in the spherical basis ($\mathbf{\hat{e}}_{\pm1}=\mp(\mathbf{\hat{x}}\pm i\mathbf{\hat{y}})/\sqrt2, \mathbf{\hat{e}}_0=\mathbf{\hat{z}}$), we write the electric dipole matrix element for an $\vert F,m_F\rangle \rightarrow \vert F',m'_F\rangle$ transition as  \cite{STE2010}:
\begin{equation}
\begin{split}
\langle F,m_F\vert er_q\vert F',m'_F\rangle=\langle J\lVert e\mathbf{r}\rVert J'\rangle(-1)^{F+J+1+I}\times\\ \sqrt{2J+1} \sqrt{2F'+1}\begin{Bmatrix} J & J' & 1 \\  F' & F & I \end{Bmatrix} \bra F,m_F|F',1,m'_F, q\ket,
\label{Eq:mat-element}
\end{split}
\end{equation}
where $q=-1,0,1$ is the spherical basis index, $\langle J\lVert e\mathbf{r}\rVert J'\rangle$ - the reduced matrix element, $J$ - the total electron angular momentum number, $I$ - the nuclear spin number, the curly brackets hold the Wigner's $6J$ symbol, and the last factor is the CGC.

Let us examine the absorption of light at frequency $f_L$ by an alkali vapor (atomic mass $m$, temperature $T$). For any $\vert F,m_F\rangle \rightarrow \vert F',m'_F\rangle$  transition (of frequency $f_{FF'}$) the light is detuned by $\Delta_{FF'}=f_{FF'}-f_L$. The light absorption rate is proportional to the square of the dipole matrix element, to the intensity of the relevant light component and to the fraction of vapor atoms having the velocity that Doppler shifts the light by $\Delta_{FF'}$. This fraction is proportional to $exp[-(\Delta_{FF'}/ \sigma_D)^2/2]$, with the Doppler standard deviation $\sigma_D=f_{FF^{'}}\cdot \sqrt { k_BT/mc^2}$.

We define the {\it relative} light absorption rate $\Gamma_{m_F}^{rel}$, as:
\begin{eqnarray}\Gamma_{m_F}^{rel}= \sum_{F'=F-1}^{F'=F+1}e^{-\frac{1}{2}\left(\frac{\Delta_{F F'}}{\sigma_D}\right)^2}(2F'+1) \begin{Bmatrix} J & J' & 1 \\  F' & F & I \end{Bmatrix}^2 \nonumber \\ \times \sum_{q=-1}^{q=1}|E_{-q}|^2  \bra F,m_F|F',1,m'_F,q\ket ^2,
\label{Eq:Gamma rel}
\end{eqnarray}
where we sum over all the possible transitions from a particular $\vert F,m_F\rangle$ sub-level.  Being only interested in the {\it relative} light absorption rates, we ignore all factors in Eq. (\ref {Eq:mat-element}) that are independent of $m_F, m'_F$ and $F'$. We also ignore the natural line width, the laser line width and the Zeeman splitting, as these are small compared to the Doppler broadening. In case the pressure shift and broadening are significant (e.g. in vapor cells containing buffer gas), one should modify the transition frequencies $f_{F F'}$ with the relevant shifts and replace the Doppler term in  Eq. (\ref{Eq:Gamma rel}) with the appropriate Voigt function \cite{ROTONDARO1997}.

The normalized polarization components $E_q$ ($|{\bf E}|^2=1$) are defined in the coordinate system where $\hat{z}$ is the direction of the magnetic field ${\bf B}$ (the quantization axis). As $E_0=E_z$ and
$E_{\pm 1}=(E_x\pm iE_y)/\sqrt{2}$, one finds for a linear polarization vector:
\begin{equation}
E_0=-\sin\theta\cos\phi;\hspace{6pt} E_{\pm 1}=(\cos\theta\cos\phi\pm i\sin\phi)/\sqrt{2}
\label{Eq:comp}
\end{equation}
where $\theta$ is the angle between ${\bf B}$ and the light wave-vector ${\bf k}$
and $\phi$ is the angle between the polarization vector ${\bf E}$ and the ${\bf B}$-${\bf k}$ plane.
Consequently, $|E_{+1}|=|E_{-1}|$ and $|E_0|^2=1-2|E_{+1}|^2$.

For any two ground state Zeeman sub-levels  $\vert F, m_{1} \rangle$, $\vert F, m_{2} \rangle$ the absorption rate of linearly polarized light will be the same if:
\begin{equation}
\Gamma_{m_{1}}^{rel}(f_L)-\Gamma_{m_{2}}^{rel}(f_L)=0.
\label{Eq:simp}
\end{equation}
Using the definition of $\Gamma_{m_F}^{rel}$ [Eq. (\ref{Eq:Gamma rel})] and the explicit expression for the CGCs, the condition of Eq. (\ref{Eq:simp}) becomes:
\begin{eqnarray}
(1-3|E_{+1}|^2)\cdot (m_{1}^2-m_{2}^2) \cdot \sum_{F'=F-1}^{F'=F+1}e^{-\frac{1}{2}\left(\frac{\Delta_{F F'}}{\sigma_D}\right)^2}\nonumber \\ \times\begin{Bmatrix} J & J' & 1 \\  F' & F & I \end{Bmatrix}^2\cdot \frac{(3\delta_{FF'}-1)(2F'+1)+F-F'}{2F'(F'+1)}=0,
\label{Eq:simp1}
\end{eqnarray}
where $\delta_{F,F'}$ is the Kronecker delta, and where for $F'=0$ the fraction in the sum equals $-1$. Let us emphasize that it is the inherent properties of the CGCs which enable to factor out the $(m_{1}^2-m_{2}^2)$ and the $(1-3|E_{+1}|^2)$ components when moving from Eqs. (\ref{Eq:Gamma rel}) and (\ref{Eq:simp}) to Eq. (\ref {Eq:simp1}).

Eq. (\ref {Eq:simp1}) has two trivial solutions corresponding to the first two factors: the first when $m_2=-m_1$, implying that always $\Gamma_{m_F}^{rel} = \Gamma_{-m_F}^{rel}$, and the second when the three polarization components have equal intensity
($|E_{+1}|^2=1/3$, or $\cos^2\theta=2/3$ when $\phi=0$), the latter being responsible for the two $\Delta\Gamma_F=0$ lines at constant $\theta$ clearly visible in Fig. 1 ($\theta=0.615$  and $\theta=\pi-0.615$). This condition is automatically valid for unpolarized light, and indeed it is well known that when all light components have equal intensity, the absorption is independent of $m_F$ \cite{STE2010}.

The third factor in Eq. (\ref {Eq:simp1}), which sums three terms, is the subject of this work.
At specific frequencies where this sum is zero, the absorption rate becomes independent of $m_F$ regardless
of the direction of polarization.
{\it Magic frequencies} always exist for every $F$ state, as the sum must have zero values for specific laser
frequencies $f_L$. To show this, we note that while the
exponent and the squared $6J$ symbol are always positive, only the $F'=F$ term is positive. The three terms are weighted by the Doppler distributions. If $f_L$ is at resonance with $F'=F$ then this term dominates and makes the sum positive, while if $f_L$ is resonant with one of the $F'=F\pm 1$ states, the sum is negative.  This is ensured by the fact that the sum vanishes if the weights on all $F'$ terms are equal,  namely, in a situation where the Doppler distribution is much broader than the hyperfine splitting of the $J'$ level. As the sum is a continuous function of $f_L$, it follows that it must vanish at least at one {\it magic frequency} between two resonances with $F'$ states.

\begin{figure}[h!] \centering \includegraphics[width=0.48\textwidth]{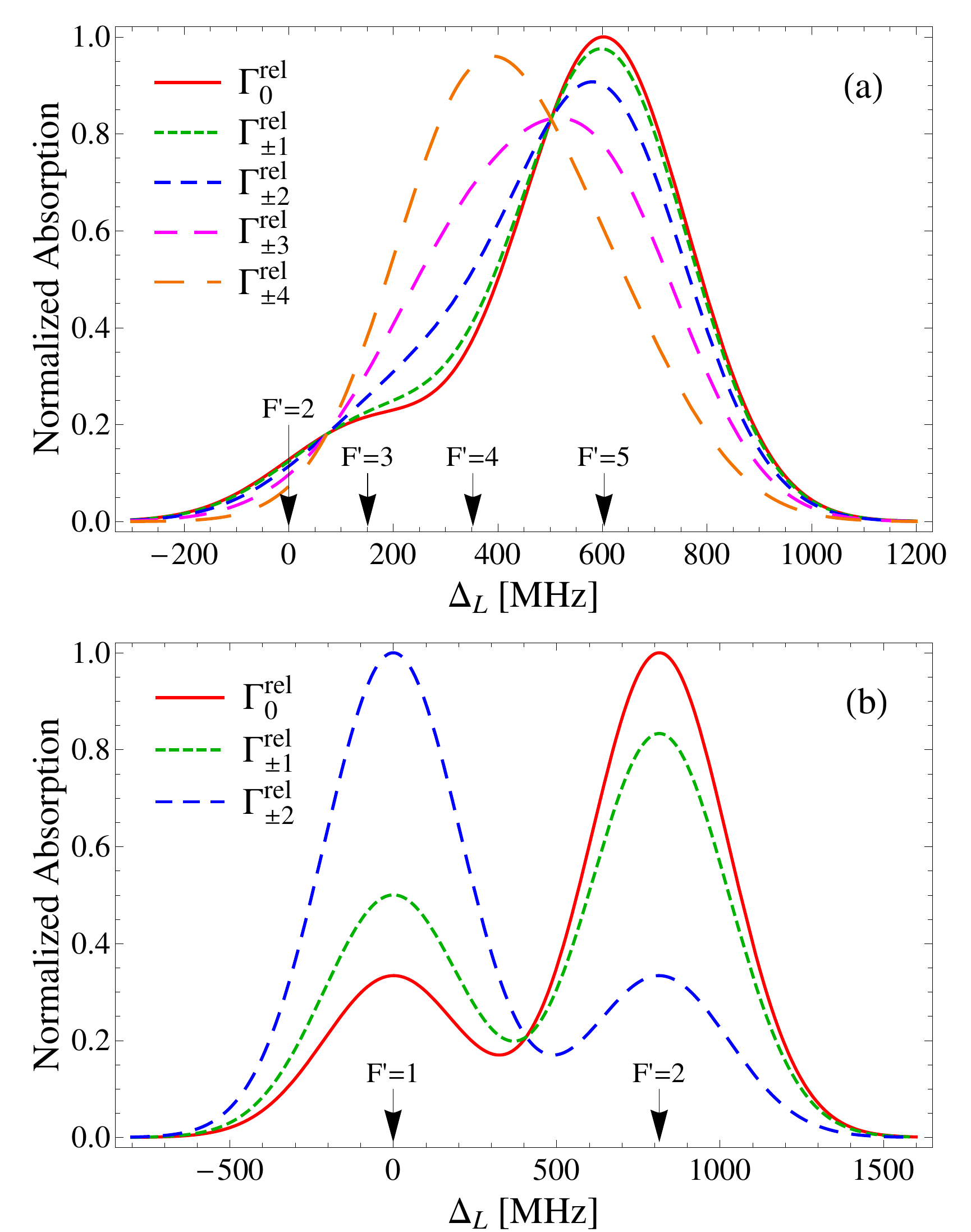}
\caption{(Color online) Relative absorption profiles [normalized $\Gamma_{m_{F}}^{rel}$, Eq. (\ref {Eq:Gamma rel})] for Zeeman sub-levels  of cesium and rubidium. At least one point where all the $\Gamma_{m_{F}}^{rel}$ are equal (i.e. $\Delta\Gamma_F=0$) always exists. These points define the {\it magic frequencies}. Shown are profiles for (a) the $6^2S_{1/2},F=4 \rightarrow 6^2P_{3/2}$ transitions of cesium ($D_2$ line), and (b) the $5^2S_{1/2},F=2 \rightarrow 5^2P_{1/2}$ transitions of $^{87}$Rb ($D_1$ line). $\Delta_L$ is the detuning of the light from the frequency difference between the ground state $|F\rangle $ and the lowest $|F' \rangle $ hyperfine state. Transition frequencies to all $F'$ levels are indicated by arrows. These absorption plots are given for linearly polarized light with $\phi=0$ (both), $\theta=\pi/2$ in (a) and $\theta=0$ in (b) [Eq. (\ref {Eq:comp})], and $T=295K$.} \label{fig:Gamma rel} \end{figure}

Fig. \ref{fig:Gamma rel} presents plots of  $\Gamma_{m_{F}}^{rel}$ vs. light frequency. Let us denote the light frequency $f_L$ by its detuning $\Delta_L$ from the frequency matching the energy difference between the hyperfine ground state $|F\rangle$ and the lowest $|F'\rangle$ state in the relevant excited hyperfine manifold. Fig. \ref{fig:Gamma rel}a shows 5 plots of $\Gamma_{m_{F}}^{rel}$ for $m_F=0...\pm 4$ of the $6^2S_{1/2},F=4 \rightarrow 6^2P_{3/2}$ transitions of cesium ($D_2$ line), interacting with linearly polarized light. Two frequencies at which all $\Gamma_{m_{F}}^{rel}$ are equal are clearly visible. These are the {\it magic frequencies}  $\Delta_L^{Magic}$. Fig. \ref{fig:Gamma rel}b contains plots for the $5^2S_{1/2},F=2 \rightarrow 5^2P_{1/2}$ transitions of $^{87}$Rb ($D_1$ line), again demonstrating the {\it magic frequency}. Similar results may be observed for $^{85}$Rb, sodium and other alkali atoms.

Finally, in order to characterize the difference between the absorption rates of light by atoms in each of the Zeeman sub-levels of a given $|F\rangle $ state, we define :
\begin{equation}
\Delta\Gamma_F\equiv \bigl[max(\Gamma_{m_F}^{rel})-min(\Gamma_{m_F}^{rel})\bigr]/S_F^M,
\label{Eq:DeltaGamma}
\end{equation}
where the max/min scan the $m_F$ space, and where
\begin{equation}
S_F^M=max(S_F), \hspace{4pt}\text {with}\hspace{4pt} S_F\equiv \frac{1}{2F+1}{\sum_{m_F=-F}^F\Gamma_{m_F}^{rel}},
\label{Eq:SF}
\end{equation}
where the max scans the frequency space.

$\Delta\Gamma_F$ is a function of $\Delta_L$, $\theta$, $\phi $ and the temperature. The quantity $S_F$,
defined to represent absorption when all the Zeeman sub-levels are equally populated, serves as a measure of the total atom-light interaction strength.
It is a function of $\Delta_L$, but is independent of both $\theta$ and $\phi$ by definition [in Eq. (\ref{Eq:SF}) all sub-levels are equally populated so there is no preferred direction]. As, at the {\it magic frequencies}, all Zeeman sub-levels contribute equally to the absorption rate, and as their sum  $S_F$ is independent of $\theta$ and $\phi$, the interaction with each sub-level must be independent of these angles as well. Thus, from a fundamental point of view, the {\it magic frequency} represents a unique cancelation effect in which light-matter interaction becomes rotationally invariant although the atomic sample as well as the light beam and its polarization all have a well defined direction. As $\rho^{(0)}$ is the only PM which is a scalar, this means that the contributions of all other PMs cancel out.

We numerically analyze $\Delta\Gamma_F$ for the $|F=2\rangle$ $D_2$ transitions of $^{87}$Rb using the surface shown in Fig. \ref{fig:3d} and find that $\Delta\Gamma_F=0$ for two {\it magic frequencies}: $\Delta_L^{Magic}=385 $ and $-318$ MHz.  For the lower frequency $S_F$ [Eq. (\ref {Eq:SF})] is very small, indicating a negligible interaction with laser light for such detuning. We also find that the value of this frequency is very sensitive to the temperature and to the type of broadening (Doppler or Voigt). Thus, it is of little practical value. On the other hand, the higher {\it magic frequency}, $\Delta_L^{Magic}=385$ MHz, is located near the maximum value of $S_F$, is only very weakly dependent on the temperature ($\Delta_L^{Magic}$ changes by $<50$ kHz/K), and for a range of about $\pm10$ MHz around  $\Delta_L^{Magic}$ we find $\Delta\Gamma_F<0.01$, indicating a nearly equal absorption rate from all the Zeeman sub-levels. We conclude that the absorption of a laser beam tuned to this {\it magic frequency}  is insensitive of $\theta$, $\phi$ and $m_F$, so that it can be used for an accurate measurement of $\rho^{(0)}$.

We have preformed an exact calculation of absorption rates based on numerical diagonalization of the full atomic Hamiltonian including the interaction with an external static magnetic field, and have found that in weak magnetic fields {\it magic frequencies} indeed exist as predicted by our simple model (the full calculation gives $\Delta\Gamma_F<0.01$ for $|\mathbf{B}|<1$ G for any set of parameters).

To experimentally demonstrate the {\it magic frequency} we  change the population distribution between the Zeeman sub-levels and show that at $\Delta_L^{Magic}$ the optical probe becomes insensitive to these changes. We use a $^{87}$Rb vapor cell with 7.5 Torr of neon buffer gas. We independently measure a pressure shift (due to the buffer gas) of ${-30\pm10}$ MHz (similar to previous measurements \cite {ROTONDARO1997}),  bringing our prediction of the {\it magic frequency} in this cell to $\Delta_L=355\pm10$ MHz.

We use a setup similar to the one used by Bhaskar \cite {Bhaskar2003}. Applying a $26$ G DC magnetic field and laser beams to the $^{87}$Rb vapor, we optically pump the vapor to the $|2,2\ket$ Zeeman sub-level. Then, utilizing an 18 MHz RF field, we induce Rabi oscillations between the $|2,2\ket$ and the $|2,1\ket$ sub-levels for a variable Rabi time $t_R$. Next, we measure the absorption due to the $\vert F=2 \rangle$ level: we turn off the RF field, adiabatically reduce the magnetic field to 1 G, turn on a strong $\pi$ polarized probe beam (tunable  in a range of $\pm 200$ MHz around the {\it magic frequency}) and record the optical density (OD) of the cell from the onset of the probe beam until the OD reaches an asymptotic value. The probe beam is strong enough to pump all the $^{87}$Rb population to the $\vert F=1 \rangle$ hyperfine state within $250\,\mu$s (short relative to the $8$ ms thermal relaxation time in the cell used here), thus bringing the OD of the cell to an asymptotic value that corresponds to zero population at the $\vert F=2 \rangle$ hyperfine state. For each Rabi time $t_R$ we calculate $\Delta_{OD}$ - the difference between OD at the onset of the probe beam and the asymptotic value of OD.  We normalize  $\Delta_{OD}$ so that for vapor in thermal equilibrium $\Delta_{OD}$=5/8.

\begin{figure}[h!] \centering \includegraphics[width=0.48\textwidth]{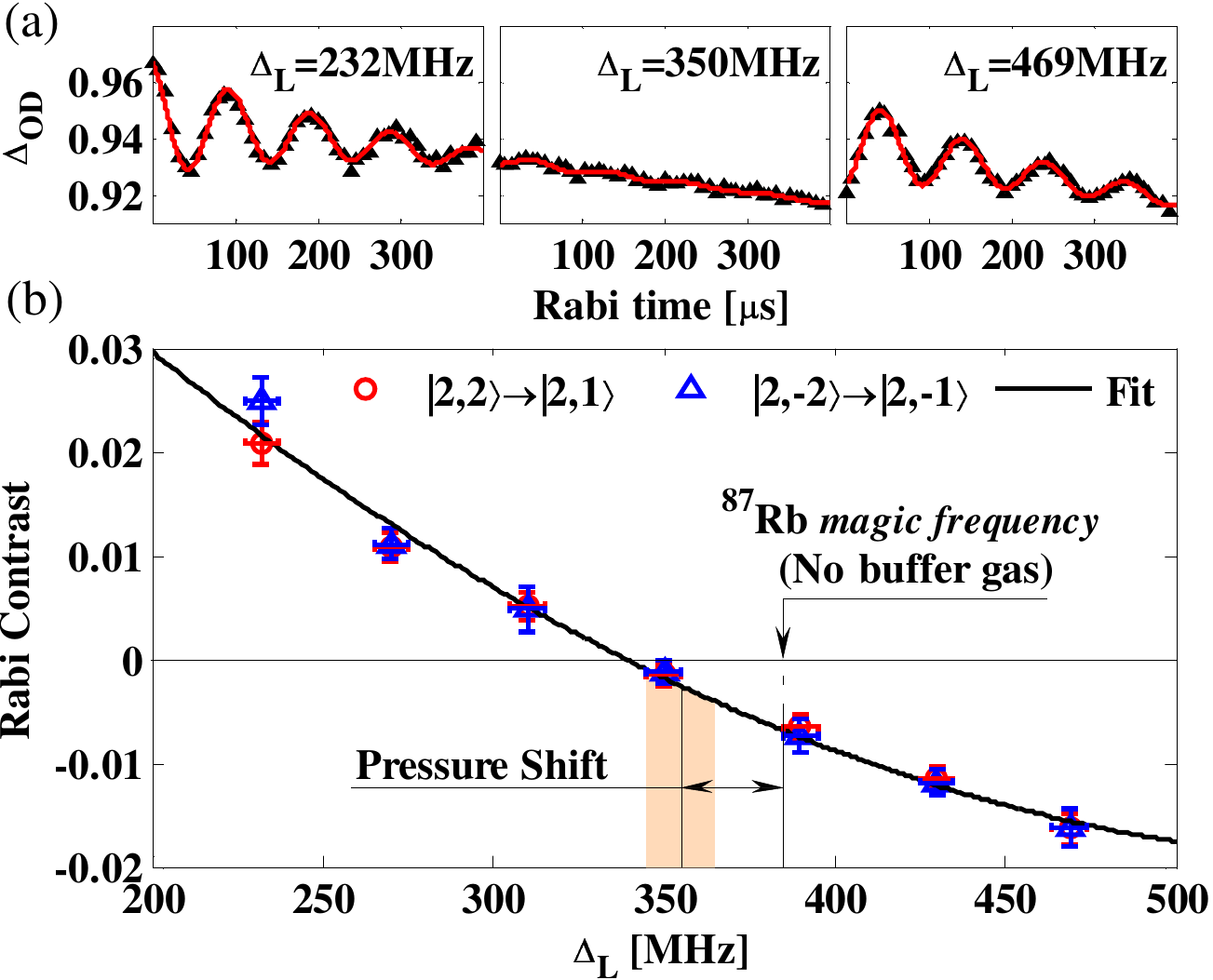}
\caption {(Color online) Experimental demonstration of the {\it magic frequency}. (a) An example of three plots of  $\Delta_{OD}$ vs. $t_R$, showing the contrast of Rabi oscillations for different frequency detunings $\Delta_L$ of the probe beam. Note the $\pi$ phase difference between the two outer plots. This is exactly as expected from the reversing of the relative absorption rates on the two sides of the {\it magic frequency}, as may be observed in Fig. \ref{fig:Gamma rel}. (b) Contrast of Rabi oscillations vs. detuning, showing that the contrast of the oscillations  between Zeeman sub-levels drops to zero as the detuning of the probe frequency is nearing 339 MHz, in good agreement with the {\it magic frequency} predicted by our model (with the pressure shift).}\label{fig:Rabi}\end{figure}

In Fig. \ref{fig:Rabi}(a) we show three example plots of  $\Delta_{OD}$ vs. $t_R$ for different frequency detuning $\Delta_L$ of the probe beam. The data points are fitted to the function $\Delta_{OD}=a+b\cdot t_R+c \cdot e^{-t_R/\tau}\cdot sin(2\pi f\cdot t_R + \Phi)$, where $c$ is the Rabi oscillation contrast. When the probe frequency detuning $\Delta_L$ is far from the {\it magic frequency}, the Rabi oscillations are clearly visible. However, when it is close to the {\it magic frequency}, the Rabi oscillations are not visible, although they do exist. Note that our measurements are sensitive enough to clearly observe Rabi contrast of less than $1\% $.

In Fig.  \ref{fig:Rabi}(b) we present the Rabi oscillations' contrast $c$ as a function of the probe beam frequency detuning $\Delta_L$, for two sets of data: pumping all the population to $\vert 2,2\rangle$, or to $\vert 2,-2\rangle$. In both cases the Rabi contrast goes to zero for a probe beam detuning of  $\Delta_L=339 \pm5$ MHz, clearly demonstrating the existence of the {\it magic frequency}. The observed value of the {\it magic frequency} is in good agreement with our simple model.

To conclude, in this Letter we present a simple model for the interaction of linearly polarized light with alkali atoms. The model reveals a {\it magic frequency} for which light is equally scattered by all the Zeeman sub-levels of the hyperfine ground state. We show analytically that such a {\it magic frequency} always exists based on the Wigner-Eckart theorem and on inherent properties of CGCs. We explore numerically the properties of the model, and use an exact calculation to determine its validity in the presence of a magnetic field.  We experimentally demonstrate the {\it magic frequency}. We expect the {\it magic frequency} to be useful in a wide range of applications, in addition to the robust measurement of the hyperfine population $\rho^{(0)}$. From a fundamental point of view, the {\it magic frequency} represents a unique cancelation effect in which light-matter interaction becomes rotationally invariant although the atomic sample as well as the light beam and its polarization all have a well defined direction. As $\rho^{(0)}$ is the only PM which is a scalar, this means that the contributions of all other PMs cancel out, a phenomenon which may shed interesting new light in the realm of group theory.

We are thankful to the members of the atom chip group and especially to Zina Binshtok and Yaniv Bar-Haim. Thanks also to the BGU machine shop team. This work was partially supported by the Israel Aerospace Industries (IAI) and the Office of the Chief Scientist in the Israeli Ministry of Industry and Trade, as part of a joint "Magneton" program.

%\bibliographystyle{apsrev4-1}
%\bibliography{Population1}

%

\end{document}